\documentclass[aps,prl,twocolumn]{revtex4-1}

\usepackage{graphicx}
\usepackage{amsmath}
\usepackage{amsfonts}
\usepackage{bm}
\usepackage{mathtools}
\usepackage{tikz}

\def\muone{{blue}}
\def\mutwo{{green}}

\def\qr{{\bf r}}                                   

\renewcommand{\vec}[1]{\bm{#1}}

\begin{document}

\title{Self-bound Bose mixtures}

\author{Clemens Staudinger$^1$, Ferran Mazzanti$^2$, and Robert E. Zillich$^1$}

\affiliation{$^1$Institute for Theoretical Physics,~Johannes Kepler University, Altenbergerstrasse 69, 4040 Linz, Austria\\
$^2$Departament de F\'{\i}sica i Enginyeria Nuclear, Campus Nord
  B4-B5, Universitat Polit\`ecnica de Catalunya, E-08034 Barcelona, Spain}

\begin{abstract}

Recent experiments confirmed that fluctuations beyond the mean-field approximation
can lead to self-bound liquid droplets of ultra-dilute binary Bose mixtures.
We proceed beyond the beyond-mean-field approximation, and study liquid
Bose mixtures using the variational hypernetted-chain Euler Lagrange method,
which accounts for correlations non-perturbatively.  Focusing on the case
of a mixture of uniform density, as realized inside large saturated droplets,
we study the conditions for stability against evaporation of one of the
components (both chemical potentials need to be negative) and against liquid-gas
phase separation (spinodal instability), the latter being accompanied by a vanishing speed
of sound.  Dilute Bose mixtures are stable only in a narrow
range near an optimal ratio $\rho_1/\rho_2$ and near the total energy minimum.
Deviations from a universal dependence on the $s$-wave
scattering lengths are significant despite the low density.

\end{abstract}


\maketitle



Ultracold quantum gases provide a rich toolbox to study correlations
in quantum many-body systems~\cite{blochRMP08} and model condensed matter physics
such as magnetic systems~\cite{micheliNaturePhys06}, solid state systems
\cite{blochNatPhys05}, or superfluidity~\cite{randeriaBCSBEC}.
A recent example is the prediction \cite{petrovPRL15} and two independent
observations~\cite{cabreraScience17,semeghini171010890} of a self-bound liquid
mixture of two ultra-dilute Bose gases ($^{39}$K atoms in two different hyperfine
states).  In this liquid state, when the attraction between different
species overcomes the single-species average repulsion, the
mean-field approach~\cite{dalfovoRMP99} would predict a collapse.
In Ref.~\cite{petrovPRL15}, correlations were taken into account
approximatively using the beyond-mean-field (BMF) approximation
\cite{larsenAnnPhys63}.  In a regime
where the BMF corrections can stabilize the binary mixture by
compensation of the mean-field attraction,
self-bound droplets are formed which live
long enough to perform measurements with the trapping potential switched off.
Being self-bound and three-dimensional, they are different from
bright solitons, which are essentially one-dimensional and have a
limited number of particles~\cite{cheineyPRL18}, while
droplets can only be formed with a critical minimum number of atoms.
On a similar footage, self-bound droplets in the region of mean-field
collapse have also been found in dipolar trapped systems of
$^{164}$Dy~\cite{Kadau_2016, FerrierBarbut_2016,schmittNature16} and $^{166}$Er~\cite{chomazPRX16}
atoms. In this case quantum fluctuations compensate the attractive
components of the dipolar interactions, as confirmed by
theory~\cite{Wachtler_2016,bombinPRL17}. Bose mixtures and dipolar
Bose gases share similarities (competition between repulsive and
attractive interactions), although the latter case is complicated
by the anisotropy of the dipolar interaction.

\begin{figure}[ht]
\begin{center}
\includegraphics*[width=0.49\textwidth]{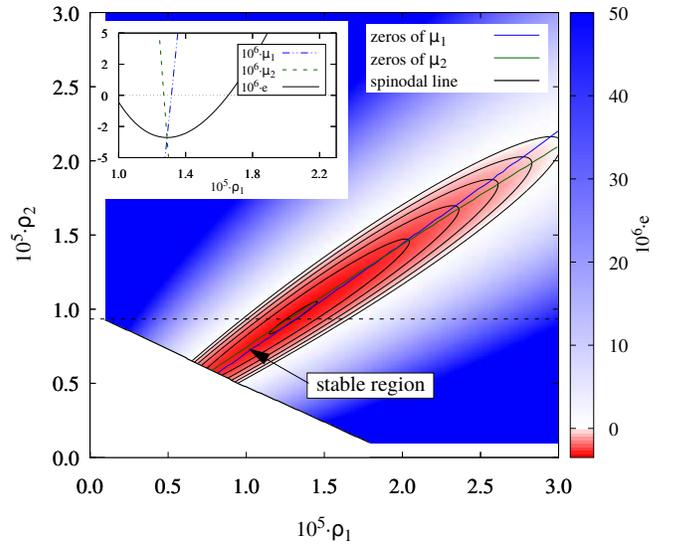}
\end{center}
\caption{
(Color online) Total energy per particle $e$
as function of $\rho_1$ and $\rho_2$, with
contour lines for energies -3.35, -2.68, -2.01, -1.34, -0.67, and 0.0
Also shown are the spinodal instability (thick black line), and the zeroes of the
chemical potentials $\mu_1$ (\muone) and $\mu_2$ (\mutwo).  Only in the narrow region
pointed at by the arrow the mixture is stable
against evaporation.  The inset shows $e$ and the chemical potentials $\mu_1$ and
$\mu_2$ along the dashed line intersecting the energy minimum.
}
\label{FIG:contour}
\end{figure}

A large enough, saturated droplet has a surface region, where the
density drops to zero, and a uniform interior, with a density plateau at
the equilibrium density $\rho_{\rm eq}$ resulting from the balance of
attractive and repulsive interactions.  In this work
we focus on the effect of self-binding rather than on the droplet surface.
Therefore we take the thermodynamic limit, $N\to\infty$ and
$V\to\infty$, with $\rho=N/V$ fixed. 
We investigate the ground state of a three-dimensional uniform
Bose mixture with partial densities $\rho_1$ and $\rho_2$ (hence a total
density $\rho=\rho_1+\rho_2$), and equal atom masses $m$.
We explore a wide range of $\rho_1$ and $\rho_2$ values, finding an
optimal ratio $\rho_1/\rho_2$ and the equilibrium density $\rho_{\rm eq}$.
We note, however, that in the presently published
experiments~\cite{cabreraScience17,semeghini171010890}, the self-bound droplets
are not saturated: they do not exhibit a central
density plateau, but an approximately Gaussian density
profile, and are so small that they are dominated
by surface effects.


The Hamiltonian of a Bose mixture is given by
\begin{equation}
H=\sum_{i,\alpha} {\hbar^2\over 2m} \Delta_{i,\alpha} 
+ {1\over 2} \sum_{\alpha,\beta}\sideset{}{'}\sum_{i,j} v_{\alpha,\beta}(|\qr_{i,\alpha}-\qr_{j,\beta}|)
\label{eq:H}
\end{equation}
where a Greek index $\alpha$ labels the component, and a Latin
index $i$ numbers the atoms of species $\alpha$.
The prime indicates that we only sum over $i\ne j$ for $\alpha=\beta$.
We use the Lennard-Jones-like interactions
$$
v_{\alpha,\beta}(r)=s_{\alpha,\beta}\left[
  \Big({\sigma_{\alpha,\beta}\over r}\Big)^{10}
- \Big({\sigma_{\alpha,\beta}\over r}\Big)^{6}
\right]\ ,
$$
with $v_{12}=v_{21}$.
The parameters of $v_{\alpha,\beta}$ are adjusted to set
the $s$-wave scattering length $a_{\alpha,\beta}$ to a desired value, which
can be done analytically\cite{padeEPJD07}. 
Since $v_{\alpha,\beta}$ has two parameters, we further
characterize $v_{\alpha,\beta}(r)$ by the effective range $r_{\alpha,\beta}^{\rm eff}$,
evaluated numerically~\cite{buendiaPRA84}.  In all calculations,
$s_{\alpha,\beta}$ and $\sigma_{\alpha,\beta}$ are chosen such that there
are no two-body bound states.

Previously, Lee-Huang-Yang corrections to the mean-field
approximation~\cite{petrovPRL15} and quantum Monte Carlo (QMC)
methods~\cite{petrovPRL16,cikojevicPRB18} have been employed.
Here we use a different approach, the variational hypernetted-chain Euler Lagrange (HNC-EL) method.
HNC-EL is computationally very economical like the BMF approximation, but has
the advantage of including correlations in a non-perturbative manner.  This leads to
a strictly real ground state energy, in contrast to the BMF approximation where the
energy of a uniform self-bound mixture has an unphysical small imaginary part.
The two-component HNC-EL method has been described in Ref.~\cite{chakrabortyPRB82}
and in a different formulation in Ref.~\cite{campbellAnnPhys72},
and has been recently generalized to multi-component Bose mixtures~\cite{hebenstreitPRA16}.
The starting point is the variational Jastrow-Feenberg
ansatz~\cite{Feenberg} for the ground state consisting of a product of pair correlation functions for
a multi-component Bose system,
\begin{equation}
  \Psi_0(\{\qr_{i,\alpha}\}) = \exp\Big[{1\over 4}\sum_{\alpha,\beta}\sideset{}{'}\sum_{i,j}
               u_{\alpha,\beta}(|\qr_{i,\alpha} - \qr_{j,\beta}|)\Big]\ .
\label{eq:Psi0}
\end{equation}
The many-body wave function $\Psi_0$ does not contain one-body functions $u_{\alpha}(\qr_{i,\alpha})$ because
we consider a uniform system.
Higher order correlations such as
triplets $u_{\alpha,\beta,\gamma}(\qr_{i,\alpha},\qr_{j,\beta},\qr_{k,\gamma})$
have been incorporated approximately for helium~\cite{fabrociniPRB84,Kro86}, but
are neglected here because their contribution is very small at low density.

We solve the Euler-Lagrange equations
$\delta e/\delta g_{\alpha,\beta}(r)=0$, where the energy per particle
$e= {1\over N}{\langle\Psi_0|H|\Psi_0\rangle\over \langle\Psi_0|\Psi_0\rangle}$
is
\begin{equation}
  e = \sum_{\alpha,\beta}{\rho_\alpha\rho_\beta\over 2\rho}\int d^3r\, g_{\alpha,\beta}(r)\big[
  v_{\alpha,\beta}(r)-{\hbar^2\over 4m}\Delta u_{\alpha,\beta}(r)
  \big]
\label{eq:e}
\end{equation}
in terms of the pair distribution function
$$
  g_{\alpha,\beta}(r) = {1+\delta_{\alpha\beta}\over \rho_\alpha\rho_\beta}
  {\delta \ln\langle\Psi_0|\Psi_0\rangle\over \delta u_{\alpha,\beta}}\ .
$$
Partial summation of the Meyer cluster diagrams for
$\ln\langle\Psi_0|\Psi_0\rangle$ in the HNC/0 approximation provides
a relation between $g_{\alpha,\beta}$ and $u_{\alpha,\beta}$ \cite{hansen,hebenstreitbachelor}.
A practical formulation of the resulting HNC-EL equations to be solved
for $g_{\alpha,\beta}$ can be found in Ref.~\cite{hebenstreitPRA16}.
From $g_{\alpha,\beta}(r)$ we can calculate the static structure functions
$S_{\alpha,\beta}(k)=\delta_{\alpha\beta}+\sqrt{\rho_\alpha\rho_\beta}\ {\rm FT}[g_{\alpha,\beta}-1]$
(FT denotes Fourier transformation), needed for the calculation of excitations.


At low densities, a uniform binary Bose mixture of two species of
equal mass is characterized by the scattering lengths $a_{11}$, $a_{12}$, and
$a_{22}$, and the partial densities $\rho_1$ and $\rho_2$.  However,
our results depend also on the next term in the expansion
of the scattering phase shift, the effective range $r_{\alpha,\beta}^{\rm eff}$~\cite{Landau3}
leading to a total of 8 parameters $\{\rho_\alpha,a_{\alpha,\beta},r_{\alpha,\beta}^{\rm eff}\}$
to characterize our uniform binary Bose mixtures.
We use $a_{11}$ as length unit and $E_0\equiv \hbar^2/m a_{11}^2$ as energy unit.
For $^{39}K$ used in experiments~\cite{cabreraScience17,semeghini171010890},
we have $a_{11}=35.2a_B$ and $E_0 = 3.55$mK.


We use the combinations of scattering lengths $a_{\alpha\beta}$
from the experiments reported in Ref.~\cite{cabreraScience17},
which are very similar to those in Ref.~\cite{semeghini171010890}.  A negative
value of $\delta a=a_{12}+\sqrt{a_{11}a_{22}}$ is necessary
for a self-bound mixture.  Before investigating the dependence on
$\delta a$, we study the dependence on the partial densities $\rho_1$ and $\rho_2$.
Fig.~\ref{FIG:contour} shows a map of the energy per particle $e$ as function of $\rho_1$ and $\rho_2$
for the experimental scattering length values corresponding to $\delta a=-5.5a_B$, which
is $\delta a=-0.156$ in our length unit $a_{11}$ and the most negative value in
Ref.~\cite{cabreraScience17}.  The other scattering lengths are
$a_{22}=1.86$ and $a_{12}=-1.52$.  The effective ranges are
$r_{11}^{\rm eff}=5.2$, $r_{12}^{\rm eff}=33.0$, and $r_{22}^{\rm eff}=43.2$.
Negative energies, where the mixture is a self-bound liquid, are shown by a red color range,
together with contour lines, positive energies by a blue color range.
Thus, as predicted by BMF calculations~\cite{petrovPRL15}
and confirmed by experiments~\cite{cabreraScience17,semeghini171010890},
we find a liquid state for $\delta a<0$.
In the phase space $(\rho_1,\rho_2)$ the self-bound states form a narrow
valley following a typical optimal ratio $\rho_1/\rho_2$.
The phase space of meaningful combinations $(\rho_1,\rho_2)$ ends at the
spinodal line (thick black line in Fig.\ref{FIG:contour}).
Approaching this line, the {\em uniform} mixture becomes
sensitive to long-wavelength density oscillations (see below). At the spinodal line
infinitesimal density fluctuations trigger a liquid-gas phase separation.

While in a uniform mixture we can choose any $\rho_1$ and $\rho_1$, a finite droplet 
adjusts its radius to minimize the energy, attaining the equilibrium
(zero pressure) density inside the droplet.  The situation is more complicated
for a mixture because the droplet radius affects only the total density,
but not necessarily the ratio $\rho_1/\rho_2$.  The latter can be adjusted
by evaporating one component or by phase separation.  Therefore
we calculate the chemical potential of component $\alpha$,
$\mu_\alpha(\rho_1,\rho_2) = e(\rho_1,\rho_2) + \rho{\partial e(\rho_1,\rho_2)\over \partial\rho_\alpha}$.
If $\mu_\alpha>0$ a particle
of species $\alpha$ is not bound to the mixture -- the energy is lowered
by removing it.  A stable
droplet requires both $e<0$ (red valley in Fig.\ref{FIG:contour}) and
$\mu_\alpha<0$.  The \muone\ line in
Fig.\ref{FIG:contour} are the zeroes of $\mu_1$, with $\mu_1<0$ {\em above} this
line.  Similarly the \mutwo\ line are the zeroes of $\mu_2$, with $\mu_2<0$
{\em below} this line.  Hence only the narrow region pointed at by the arrow
is stable against evaporation; this region includes of course the equilibrium
energy $e_{\rm eq}=\min[e]$.
The inset of Fig.\ref{FIG:contour}
shows $e$, $\mu_1$ and $\mu_2$ along the dashed line as function of $\rho_1$ for
a fixed value $\rho_2 a_{11}^3 = 0.934$, such that we intersect the
equilibrium energy.  $\mu_\alpha$ is very sensitive
to the partial density, which explains why the region where
both $\mu_\alpha<0$ is so narrow.
If a droplet is prepared outside the stable region,
particles evaporate and the system moves on the energy surface
until it is stable.

Our results for
$e$ and $\mu_\alpha$ mean that large droplets will reach the equilibrium
energy $e_{\rm eq}$ by a combination of evaporating superfluous
particles and adjusting the droplet radius.  In the case discussed so
far, $\delta a=-0.156$, the density ratio at the equilibrium
energy predicted from our HNC-EL results in Fig.\ref{FIG:contour})
is $\rho_1/\rho_2=1.380$, which is to be compared with
the optimal mean-field ratio~\cite{PethickSmith} $\rho_1/\rho_2=\sqrt{a_{22}/a_{11}}=1.363$.
The latter is a very good approximation even though
the mean-field approximation does not even predict a liquid state.
As seen in the inset of Fig.\ref{FIG:contour}, $e$ changes
very little if the density ratio is slightly changed; therefore
for further calculations of $e$ we use the mean-field ratio.


\begin{figure}[ht]
\begin{center}
\includegraphics*[width=0.50\textwidth]{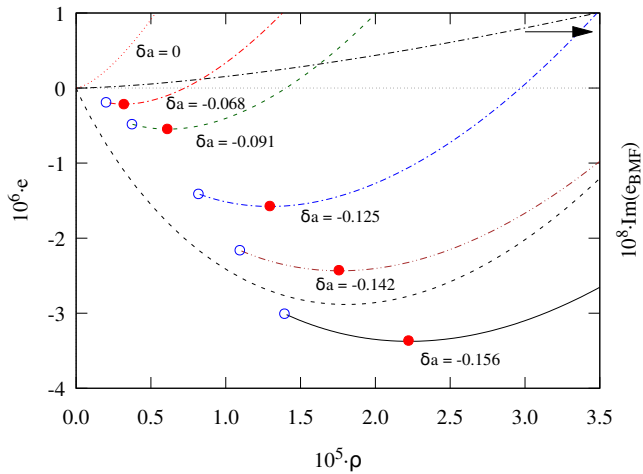}
\end{center}
\caption{
(Color online) Energy per particle $e$ as function of the total density $\rho$
for several values $\delta a$.  Closed circles denote the equilibrium
density $\rho_{\rm eq}$ and energy $e_{\rm eq}$; open circles denote spinodal points.
The black dashed and dash-dotted lines are the real and imaginary parts, the latter
using a different scale, of the BMF energy for $\delta a=-0.156$.}
\label{FIG:delta}
\end{figure}

\begin{table}
\begin{tabular}{|c|c|c|c|c|c|c|}
\hline
$\delta a [a_{B}]$ & $\delta a$ & $a_{22}$ & $r_{22}^{\rm eff}$ & $\rho_1/\rho_2$ & $10^6 e_{\rm eq}$ & $10^5 \rho_{\rm eq}$ \\
\hline
$-5.5$ & $-0.156$ & $1.86$ & $43.2$ & $1.363$ & $-3.364$ & $2.221$ \\
$-5.0$ & $-0.142$ & $1.90$ & $40.3$ & $1.377$ & $-2.426$ & $1.756$ \\
$-4.4$ & $-0.125$ & $1.94$ & $37.0$ & $1.394$ & $-1.571$ & $1.294$ \\
$-3.2$ & $-0.091$ & $2.04$ & $31.2$ & $1.428$ & $-0.544$ & $0.609$ \\
$-2.4$ & $-0.068$ & $2.10$ & $27.7$ & $1.450$ & $-0.214$ & $0.319$ \\
$0.0$ & $0.0$ & $2.31$ & $19.1$ & $1.519$ & -- & -- \\
\hline
\end{tabular}
\caption{
Values for $\delta a$, $a_{22}$, $r_{22}^{\rm eff}$, and $\rho_1/\rho_2$ used to obtain
the results shown in Fig.~\ref{FIG:delta}, as well as the equilibrium energy and density
obtained from these results.
Lengths and energies are in units of $a_{11}$ and $E_0$ (see text) if not otherwise stated.
}\label{tab:2}
\end{table}

When $\delta a$ increases towards zero, the liquid becomes less bound,
until it is no longer self-bound at $\delta a=0$.
In Fig.~\ref{FIG:delta} the energy per particle $e$ as function of total
density $\rho$ is shown for several values of $\delta a$ in the
range $[-0.156,0]$, corresponding to the
range of values in experiments~\cite{cabreraScience17,semeghini171010890}
where $\delta a$ is adjusted by changing the $s$-wave scattering length
$a_{22}$ via a magnetic field.  We follow this protocol
and modify the strength $s_{22}$ of $v_{22}(r)$ to
obtain the corresponding $a_{22}$, which also changes $r_{22}^{\rm eff}$;
$v_{11}$ and $v_{12}$ are not changed and chosen as above.
Table~\ref{tab:2} lists the values of $a_{22}$, $r_{22}^{\rm eff}$, and $\rho_1/\rho_2$
for Fig.~\ref{FIG:delta}.
The equilibrium energies $e_{\rm eq}$ and densities $\rho_{\rm eq}$ are
marked by filled circles, and
are also listed in table~\ref{tab:2}.  Naturally, both $e_{\rm eq}\to 0$
and $\rho_{\rm eq}\to 0$ as $\delta a\to 0$. For $\delta a=0$, $e>0$ and
the mixture is not liquid anymore.  The spinodal densities, where a uniform
liquid becomes unstable against infinitesimal density fluctuations, are marked
by open circles in Fig.~\ref{FIG:delta}.
Also shown in Fig.~\ref{FIG:delta} is the energy per particle $e_{\rm BMF}$
calculated in the BMF approximation~\cite{petrovPRL15}
for $\delta a=-0.156$.
Since $e_{\rm BMF}$ is complex, we show both the real and the small imaginary
part of $e_{\rm BMF}$ (note the different energy scale for the latter).
The BMF approximation fails to predict the spinodal instability and
$e_{\rm BMF}$ extends all the way to $\rho=0$ 

\begin{figure}[ht]
\begin{center}
\includegraphics*[width=0.48\textwidth]{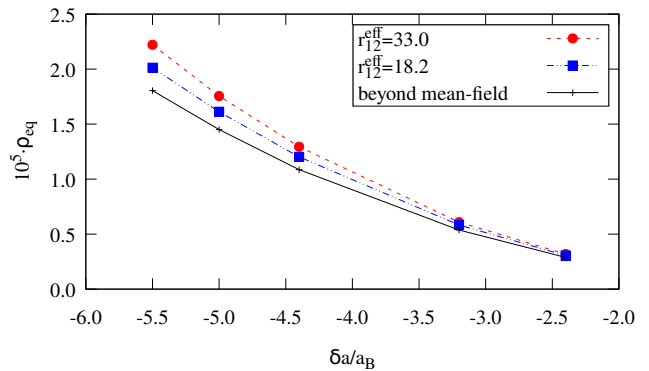}
\end{center}
\caption{
(Color online) Equilibrium density $\rho_{\rm eq}$ of the uniform Bose mixture
as function of $\delta a$, varied by changing $a_{22}$, see table~\ref{tab:2}.
For all curves, $a_{12}=-1.52$.  The circles and squares are the present HNC-EL results
obtained for $r_{12}^{\rm eff}=33.0$ and $18.2$, respectively. The line is the
BMF result.}
\label{FIG:rhoeq}
\end{figure}

The density is more accessible to measurement than the energy, e.g.\
in Refs.~\cite{cabreraScience17,semeghini171010890} the central density of
droplets was measured.  In Fig.\ref{FIG:rhoeq} we summarize the results shown
in Fig.\ref{FIG:delta} by plotting the equilibrium total density $\rho_{\rm eq}$
as a function of $\delta a$ (filled circles).
Also shown in Fig.\ref{FIG:rhoeq} is the BMF result for $\rho_{\rm eq}(\delta a)$,
obtained as the minimum of the real part of the BMF energy,
which qualitatively agrees with HNC-EL, but predicts a somewhat lower equilibrium
density.
The square symbols in Fig.\ref{FIG:rhoeq} show our results
for $\rho_{\rm eq}(\delta a)$, if we choose different parameters $s_{12}$
and $\sigma_{12}$ in $v_{12}(r)$ such that we keep $a_{12}=-1.519$, while
changing the effective range from $r_{12}^{\rm eff}=33.0$ (upper curve)
to $r_{12}^{\rm eff}=18.2$ (lower curve).  This demonstrates that the
results are not universal; they depend not only
the $s$-wave scattering lengths, but at least also on the effective ranges.

\begin{figure}[ht]
\begin{center}
\includegraphics*[width=0.48\textwidth]{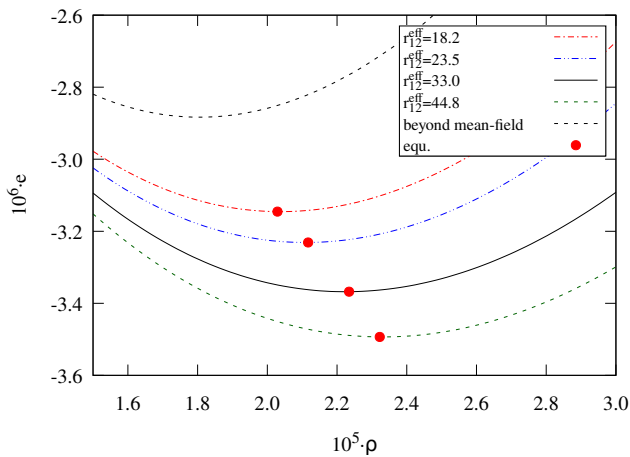}
\end{center}
\caption{
(Color online) $e(\rho)$ for $\delta a=-0.156$
and four models for the inter-species interaction $v_{12}$
corresponding to effective ranges $r_{12}^{\rm eff}=18.2; 23.0; 33.0; 44.8$.
Also shown is the BMF result ${\rm Re}[e_{\rm BMF}]$.
}
\label{FIG:reff}
\end{figure}

In Fig.\ref{FIG:reff} shows the dependence of
the energy per particle $e(\rho)$ on $r_{12}^{\rm eff}$ for $\delta a=-0.156$,
with $a_{12}=-1.519$ and the other scattering lengths as above,
for different values $r_{12}^{\rm eff}=18.2; 23.0; 33.0; 44.8$.
The dependence on $r_{12}^{\rm eff}$ is significant, with $e$ varying by $10\%$
and $\rho_{\rm eq}$ varying by $13\%$ for this range of $r_{12}^{\rm eff}$ values.
The BMF energy ${\rm Re}[e_{\rm BMF}]$ agrees better with HNC-EL for smaller $r_{12}^{\rm eff}$.
Alkali interaction potentials have a finite effective range
that is often much larger than the $s$-wave scattering
length, see e.g.\ table~1 in Ref~\cite{flambaumPRA99}.
Considering the low equilibrium densities $\rho_{\rm eq}$, it might
appear surprising to find this non-universal behavior.  We note, however,
that for small $\delta a$ the mean-field energy is the result of large
cancellations of negative and positive contributions.
Therefore it is plausible that a dependence on higher-order parameters such
as the effective range becomes visible.

\begin{figure}[ht]
\begin{center}

\includegraphics*[width=0.48\textwidth]{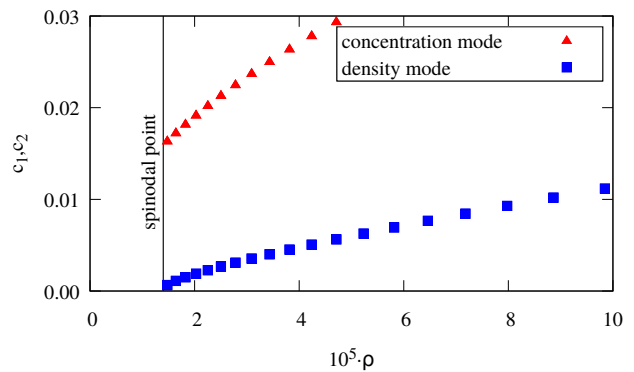}
\end{center}
\caption{
Long-wavelength phase velocities in the two-component Bijl-Feynman approximation
for density oscillations (lower curve) and concentration oscillations (upper curve)
as function of total density.  The vertical line denotes the spinodal instability.}
\label{FIG:stucturespeed}
\end{figure}

The spinodal instability (thick black line in Fig.~\ref{FIG:contour})
can be relevant for the preparation
of the liquid droplets, achieved by ramping one of the scattering lengths.
During a fast ramp, the mixture may visit the
``forbidden'' region of the $(\rho_1,\rho_2)$-phase space and
can condense into multiple droplets.  To characterize the uniform liquid mixture
near this instability in more detail,
we choose $a_{\alpha,\beta}$ and $r_{\alpha,\beta}^{\rm eff}$ as for Fig.~\ref{FIG:contour}
corresponding to $\delta a=-0.156$, and the mean-field
optimal ratio $\rho_1/\rho_2=1.363$.
A simple approximation for the excitation spectrum of a Bose mixture is given
by the Bijl-Feynman approximation\cite{Feynman3}, which provides
a good estimate of the long wave length dispersion.
A mixture supports density and concentration
oscillations, with dispersion relations $\epsilon_1(k)$ and
$\epsilon_2(k)$, respectively. They can be easily calculated from the static structure
functions $S_{\alpha\beta}(k)$
by solving the
eigenvalue problem ${\hbar^2k^2\over 2m} \vec\psi_i=\epsilon_i(k) {\bf S}(k) \vec\psi_i$
where ${\bf S}$ is the $2\times 2$ matrix with elements $S_{\alpha\beta}(k)$ 
and $\vec\psi_i$ are 2-component vectors.  Fig.~\ref{FIG:stucturespeed} shows
the long-wavelength phase velocities $c_i=\lim_{k\to 0}d\epsilon_i(k)/dk$ for the
density and concentration mode.  The density mode has lower energy
than the concentration mode for all densities shown in
Fig.~\ref{FIG:stucturespeed}, including the equilibrium density.  While
$c_2$ is finite and hence the mixture is stable against demixing, the
density mode becomes soft for $k\to0$ as $\rho$ is lowered, evidenced
by the vanishing speed of sound $c_1$ at the spinodal instability (vertical line),
where phase separation into liquid and gas occurs.
This is similar to the onset of the modulational instability in
dipolar Bose gases\cite{kadauNature16,Ferrier-Barbut_2016}, which however
is triggered by a vanishing roton energy with a {\em finite}
wave number~\cite{ferrierBarbutPRA18,chomazNatPhys18}.

In summary,
we analyze the properties of a liquid, i.e.\ self-bound, uniform Bose mixture
using $s$-wave scattering lengths as
in Refs.~\cite{cabreraScience17,semeghini171010890}.  With
the HNC-EL method, which includes pair correlations non-perturbatively,
we find a narrow regime of partial densities $\rho_\alpha$ where the conditions for a stable
liquid mixture are met: the energy per particle and
both chemical potentials $\mu_\alpha(\rho_1,\rho_2)$ are negative.
If $\mu_\alpha>0$, atoms of component $\alpha$ evaporate
until reaching either equilibrium or the spinodal line.
Despite their ultra-low density, the properties of these liquids
depend also on the effective ranges
$r_{\alpha,\beta}^{\rm eff}$.  This deviation from universality
was not observed in two-dimensional liquids~\cite{petrovPRL16}.
Comparison of the energies and equilibrium densities between the BMF
approximation and our HNC-EL calculations shows that the difference
can probably be attributed to the neglect of the effective range in
the BMF approximation.
Unlike BMF energies, HNC-EL energies do not have an unphysical
imaginary part for $\delta a<0$, as HNC-EL is not based on an expansion
about the (unstable) mean-field result.  We find that
the liquid can have a spinodal instability, where the speed of sound
vanishes and infinitesimal density
fluctuations lead to a separation into a liquid and gas phase.
This can be relevant during a nonadiabatic evolution of a droplet in experiments.
In the presently available experiments~\cite{cabreraScience17,semeghini171010890}
the liquid droplets are far from saturation, as evidenced by the
Gaussian shaped density profiles, and possibly not in equilibrium.
Describing small unsaturated droplets will require 
an inhomogeneous generalization of HNC-EL based on the
energy functional (\ref{eq:e}).

We acknowledge discussions with Leticia Tarruell and
financial support by the Austrian Science Fund FWF (grant No.\ P23535-N20).

\bibliography{bec,my,Mazz_Refs}
\end{document}